\documentclass[aps,
 prl,
 amssymb,showpacs,
 twocolumn
 ]{revtex4}

\usepackage{graphicx}
\usepackage{amsmath}

\begin{document}

\title{Spatial quantum correlations in multiple scattered light}

\date{\today}
\author{P. Lodahl} \altaffiliation[Present address: ]{Research Center COM, Technical
University of Denmark, Dk-2800 Lyngby, Denmark}
\email{pel@com.dtu.dk}

\author{A.P. Mosk}
\author{A. Lagendijk} \altaffiliation[Present address: ]{FOM Institute for Atomic and
Molecular Physics (AMOLF), P.O. Box 41883, 1009 DB Amsterdam, The
Netherlands. }

\affiliation{Complex Photonic Systems, Department of Science and
Technology and MESA$^+$ research institute, University of Twente,
P.O. Box 217, 7500 AE Enschede, The Netherlands.}


\begin{abstract}
We predict a new spatial quantum correlation in light propagating
through a multiple scattering random medium. The correlation
depends on the quantum state of the light illuminating the medium,
is infinite range, and dominates over classical mesoscopic
intensity correlations. The spatial quantum correlation is
revealed in the quantum fluctuations of the total transmission or
reflection through the sample and should be readily observable
experimentally.
\end{abstract}

 \maketitle

When light propagates through a disordered scattering medium, a
strongly modulated interference structure, known as a speckle
pattern, is generated. Speckle patterns are the most direct
manifestation of wave coherence in transport of light through
samples that are thicker than the transport mean free path $\ell$,
which is the average distance over which the direction of light is
diffused due to random scattering. By analyzing the statistical
properties, such volume speckle patterns reveal strong
correlations that are responsible for fundamental physical
phenomena as the memory effect \cite{Feng88,Freund88} and enhanced
mesoscopic fluctuations
\cite{Albada90,Scheffold98,Shapiro99,Martin02,Skipetrov04}.
Furthermore, clear signatures of Anderson localization of light
have been observed by analyzing intensity fluctuations
\cite{Chabanov00}.

While spatial correlations of the \emph{intensity} of multiply
scattered light have been investigated intensely, spatial
correlations hidden in the \emph{quantum fluctuations} of light
have to our knowledge not been addressed. Here we predict that
strong quantum correlations exist between spatially separated
parts of a far-field speckle pattern that have no classical
analog. The quantum correlation is infinite range, since it is
independent of the angular separation between the parts of the
speckle pattern under investigation, and dominates over mesoscopic
fluctuations. Therefore speckle patterns generated with quantum
light are much stronger correlated than classical theory predicts,
and the quantum corrections should be observable even for moderate
amount of scattering. In contrast, the magnitude of classical
intensity correlations are all found to scale with the inverse of
the mesoscopic conductance $g = N \ell/L$ where $N$ is the number
of conducting modes and $L$ is the thickness of the scattering
medium \cite{Albada90,Scheffold98,Shapiro99,Martin02,Skipetrov04}.
Only for extremely strong scattering, i.e. close to the Anderson
localization transition $(g \simeq 1)$, do such mesoscopic
correlations dominate.

The quantum correlation will be shown to depend on the quantum
state of light illuminating the multiple scattering sample. The
physical origin of the effect is due to distribution of photon
correlations of the input state over the spatial degrees of
freedom of the speckle pattern. This mechanism was previously
predicted to lead to new long-range correlations of thermal
radiation from a disordered waveguide \cite{Patra&Beenakker99}. We
evaluate the correlation function for three different quantum
states: coherent state, thermal state, and Fock state. For a
coherent state the spatial correlation vanishes, while strong
correlations and anti-correlations are found for the thermal state
and Fock state, respectively. We show that the spatial quantum
correlation can be measured conveniently by recording the quantum
noise of the total transmission or reflection from the medium.
Such measurements were carried out very recently for a coherent
state \cite{Lodahl05}, and appear feasible also in the near future
for other quantum states. Our work is connected to the field of
quantum imaging where spatial correlations are employed for
reconstructing images \cite{Lugiato02}.

\begin{figure}[t]
\includegraphics[width=0.8\columnwidth]{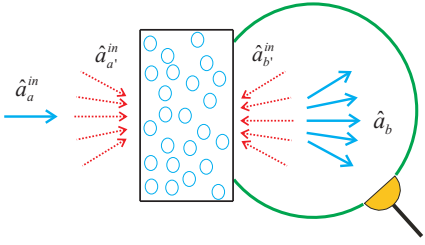}
 \caption{(color online). Total transmission through a multiple scattering sample. A quantum state of light is coupled through mode $a$ and the
 total number of photons is collected with an integrating sphere. Vacuum fluctuations in all open channels $a'\neq a$ and
 $b'$ must be included in the quantum description.  }
 \label{input-output}
\end{figure}

We describe the propagation of quantum states of light through a
non-absorbing, multiple scattering medium using the formalism
developed by Beenakker \emph{et al.} \cite{Beenakker98,Patra00}.
The model describes effectively a quasi-1D configuration, but this
approximation is known to be excellent also for a 3D slab
geometry. The annihilation operator $\hat{a}_{b}$ for an output
mode $b$ is coupled to the annihilation operators associated with
all input modes $\hat{a}_{a'}^{in}$ and $\hat{a}_{b'}^{in}$
 through the relation
$\hat{a}_{b} = \sum_{a'} t_{a'b} \hat{a}_{a'}^{in} + \sum_{b'}
r_{b' b} \hat{a}_{b'}^{in},$ where $t$ and $r$ are electric field
transmission and reflection coefficients, respectively, and the
summations are over all $N$ input and output modes. We use the
notation that $a$ denotes modes to the left and $b$ modes to the
right of the medium, c.f. Fig. \ref{input-output}. The modes
correspond to different propagation directions (k vectors) of the
incoming and outgoing light as measured directly in the far-field.

In the following, we explicitly outline the calculation of the
quantum fluctuations of the light transmitted through the sample
when light is coupled through a single input mode $a$. In the
quantum description vacuum fluctuations coupled into all other
input modes $a' \neq a$ and $b'$ must be included. The operator
$\hat{n}_b = \hat{a}_b^{\dagger} \hat{a}_b$ describes the number
of photons transmitted to the output mode $b$, and the total
number of transmitted photons is obtained by adding all output
modes: $\hat{n}_T = \sum_b \hat{n}_b.$ In an experiment, the total
transmission can be measured with an integrating sphere, cf. Fig.
\ref{input-output}. The quantum fluctuations are quantified
through the variance of the photon number:
 $\Delta n^2 \equiv \left< \hat{n}^{2} \right> - \left<
\hat{n} \right>^2,$ where $\left< \cdot \right>$ is the quantum
mechanical expectation value. The fluctuations of the total
transmission are straightforwardly found to be
\begin{equation} \Delta{n}_T^2 = \sum_{b} \Delta n_{b}^2 + \sum_{b_0} \sum_{b_1
\neq b_0} \left< \hat{n}_{b_0} \hat{n}_{b_1} \right> - \left<
\hat{n}_{b_0} \right> \left< \hat{n}_{b_1} \right>, \label{var-nT}
\end{equation}
where an index specifying the fixed input mode $(a)$ has been
omitted for brevity of notation. We immediately see that the
variance of the total transmission is different from the the sum
of the variances of all outgoing modes. Also cross-correlations
between different output modes contribute, and these correlations
are shown to be significant in the following.

In order to evaluate Eq. (\ref{var-nT}), $\left<\hat{n}_b \right>$
and $\left<\hat{n}_{b_0} \hat{n}_{b_1} \right>$ must be computed.
The non-vanishing contributions are
\begin{widetext}
\begin{subequations} \begin{eqnarray} \left<\hat{n}_b\right> &=& \sum_{a',a''}
t_{a'b}^* t_{a''b} \left<  \hat{a}{{}_{a'}^{in}}^{\dagger}
\hat{a}_{a''}^{in}  \right>, \label{av-nb} \\
\left<\hat{n}_{b_0} \hat{n}_{b_1}\right> &=&
\sum_{a',a'',a''',a''''} t_{a'b_0}^* t_{a''b_0} t_{a'''b_1}^*
t_{a''''b_1} \left<
 \hat{a}{{{}_{a'}^{in}}^{\dagger}}
\hat{a}_{a''}^{in} \hat{a}{{}_{a'''}^{in}}^{\dagger}
\hat{a}_{a''''}^{in} \right>  + \nonumber \\
&& \sum_{a',b',b'',a''} t_{a'b_0}^* r_{b'b_0} r_{b''b_1}^*
t_{a''b_1} \left< \hat{a}{{}_{a'}^{in}}^{\dagger}
\hat{a}_{b'}^{in} \hat{a}{{}_{b''}^{in}}^{\dagger}
\hat{a}_{a''}^{in} \right>.
\end{eqnarray} \label{direct+cross-corr}
\end{subequations}
 The expectation values of the various operator products can be evaluated using that
vacuum is coupled through all other channels than $a$. We obtain
\begin{subequations} \begin{eqnarray}  \left<  \hat{a}{{}_{a'}^{in}}^{\dagger}
\hat{a}_{a''}^{in}  \right> &=&  \left< \hat{n}_{a}^{in}\right> \delta_{a',a} \delta_{a'',a}, \\
 \left< \hat{a}{{}_{a'}^{in}}^{\dagger} \hat{a}_{b'}^{in}
\hat{a}{{}_{b''}^{in}}^{\dagger} \hat{a}_{a''}^{in} \right> &=&
\left< \hat{n}_{a}^{in}\right>
\delta_{a',a} \delta_{a'',a} \delta_{b',b''}, \\
 \left< \hat{a}{{}_{a'}^{in}}^{\dagger} \hat{a}_{a''}^{in}
\hat{a}{{}_{a'''}^{in}}^{\dagger} \hat{a}_{a''''}^{in} \right> &=&
\delta_{a',a} \delta_{a'''',a} \left[ \left<
\hat{n}_{a}^{in}\right> \delta_{a'',a'''} + \left(
\left<\hat{n}{_a^{in}}^2 \right> - \left< \hat{n}_a^{in}\right>
\right) \delta_{a'',a} \delta_{a''',a} \right].   \end{eqnarray}
\label{expec-values} \end{subequations}
\end{widetext}
By combining Eqs. (\ref{direct+cross-corr}) and
(\ref{expec-values}) using an identity derived from commutation
relations \cite{commutation relations}, the photon number variance
of mode $b$ and the cross-correlation between two different
directions $b_0 \neq b_1$ can be found
\begin{subequations} \begin{eqnarray} && \frac{\overline{\Delta
n_{b}^2}}{\left<\hat{n}_a^{in} \right>}= \overline{T_{ab}}
 +  \overline{T_{ab}}^2
\left(F_a^{in}
- 1 \right)  \left(2+ 8/3g \right),  \\
&& \nonumber \\ && \frac{\overline{\left< \hat{n}_{b_0}
\hat{n}_{b_1} \right>} - \overline{\left< \hat{n}_{b_0} \right>
\left< \hat{n}_{b_1} \right>}}{\left<\hat{n}_a^{in} \right>} =
\overline{T_{ab}}^2
 \left(F_a^{in} - 1 \right) \left(1+
4/3g \right). \nonumber  \\
&&
 \end{eqnarray} \label{nb-nb0nb1} \end{subequations}
Here we have introduced the intensity transmission coefficients
$T_{ab} = \left|t_{ab} \right|^2$ that have been averaged over all
realizations of disorder (denoted by a bar). Products of
transmission coefficients are expanded to first-order in $1/g$,
i.e. $\overline{T_{ab_0} T_{ab_1}} \approx \left(1+
\delta_{b_0,b_1} \right)(1 + 4/3g ) \overline{T_{ab}}^2$
\cite{Berkovits94}. The Fano factor $F_a^{in} = \Delta
(n_a^{in})^2/\left<\hat{n}_a^{in} \right>$ measures the variance
of the photon number relative to the average number of photons,
and is equal to unity for Poissonian photon statistics. We
evaluate the Fano factor for three different single-mode quantum
states of light: a coherent state (CS), a thermal state (TS), and
an n-photon Fock state (FS), which correspond to $\left(F_a^{in}
\right)_{CS}=1,$ $\left(F_a^{in} \right)_{TS}=1 +
\left<\hat{n}_a^{in} \right> $ and $\left(F_a^{in}
\right)_{FS}=0$, respectively \cite{Mandel&Wolf}.

The classical transmission coefficient can be calculated from the
theory of light diffusion, which leads to $\overline{T_{ab}} =
\ell/N L$ in the absence of absorption \cite{Rossum99}. Inserting
Eqs. (\ref{nb-nb0nb1}) in Eq. (\ref{var-nT}), we arrive at the
final expression for the quantum fluctuations of the total
transmission
\begin{equation} \frac{\overline{\Delta {n}_T^2}}{\left<\hat{n}_a^{in}
\right>}= \frac{\ell}{L} + \frac{\ell^2}{L^2} \left(F_a^{in}-1
\right) \left[ 1 + \frac{1}{g} \left(\frac{4}{3} + \frac{\ell}{L}
\right)\right] . \label{Total-Fluc} \end{equation}
Neglecting at present correction terms of order $1/g$, the
variance of the total transmission is seen to contain a linear and
a quadratic term in $\ell/L.$ The former is the sum of the
variances of each individual channel, which corresponds to
incoherently adding the fluctuations of each output mode. The
quadratic term is due to correlations between different output
modes. Note that while a quadratic scaling was found also for
transmission of classical noise through a multiple scattering
medium \cite{Lodahl05}, the contribution discussed here is a
quantum effect. The fluctuations of the total reflection can be
calculated in a similar way, and neglecting correction terms of
order $1/g$ leads to
\begin{equation} \frac{\overline{\Delta {n}_R^2}}{\left<\hat{n}_a^{in}
\right>}= \left(1 - \frac{\ell}{L} \right) + \left(1-
\frac{\ell}{L} \right)^2 \left(F_a^{in}-1 \right)  .
\label{Total-Fluc-Reflec} \end{equation}

\begin{figure}[t]
\includegraphics[width=0.8\columnwidth]{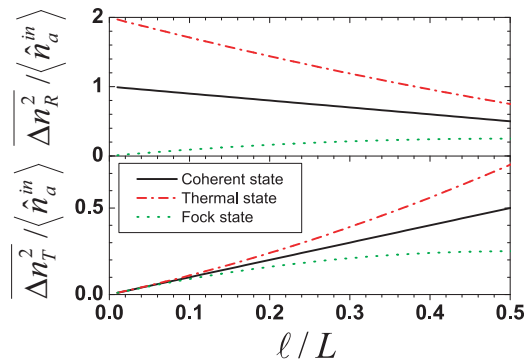}
 \caption{(color online). Variance of the quantum fluctuations of the total reflection (upper plot) and total transmission
 (lower plot) for a coherent state (full line), a thermal
 state with $\left<\hat{n}_a^{in} \right> =
 1$ (dash-dotted line) and an arbitrary Fock state (dotted line) as a
function of $\ell/L.$ All correction terms of order $1/g$ have
 been neglected. In the reflection (transmission) pronounced variations in the photon statistics are found for decreasing (increasing)
  $\ell/L$.  }
 \label{var-of-total-transmission}
\end{figure}

 Figure
\ref{var-of-total-transmission} displays the variance of the total
transmission and reflection for three different quantum states of
light as a function of the ratio of the mean free path $\ell$ to
the sample length $L$. For a coherent state, the variance scales
linearly, which was experimentally confirmed in \cite{Lodahl05}.
For other quantum states of light, a quadratic term contributes by
either a positive (for the thermal state) or negative amount (for
the Fock state), which means that the photon statistics of the
total transmission and reflection is super-Poissonian or
sub-Poissonian, respectively. These quantum correlations are found
to be most pronounced in the reflection for thick samples $(\ell/L
\rightarrow 0)$ and in the transmission for thin samples $(\ell/L
\rightarrow 1).$ This behavior can be understood intuitively since
for thick (thin) samples relatively few photons are transmitted
(reflected) and vacuum fluctuations coupled through all open modes
lead to Poissonian photon statistics. The validity of the
diffusion model for light propagation requires $\ell/L \lesssim
1$, and in this limit each transmitted or reflected photon has
experienced numerous scattering events. Remarkably the quantum
correlations of these multiply scattered photons are preserved,
and for $\ell/L \rightarrow 0$, the fluctuations of the total
reflection equal the fluctuations of the input light. This clearly
demonstrates that deep in the multiple scattering regime, the
quantum corrections are most pronounced in the reflection, and
thereby devises a route for experimental observation of quantum
correlations in a volume speckle pattern.

The quadratic terms in Eqs. (\ref{Total-Fluc}) and
(\ref{Total-Fluc-Reflec}) originate from spatial correlations
between different output directions in the speckle pattern. We
define the two-point spatial correlation function, which is
evaluated using Eqs. (\ref{direct+cross-corr}),
(\ref{expec-values}) and \cite{commutation relations}
\begin{equation} C_{b_0b_1} \equiv \frac{\overline{\left<\hat{n}_{b_0}
\hat{n}_{b_1} \right>}}{\overline{\left<\hat{n}_{b_0}\right>
\left<\hat{n}_{b_1} \right>}} = 1 +
\frac{F_a^{in}-1}{\left<\hat{n}_a^{in} \right>},
\label{two-point-transmission} \end{equation}
where $b_0 \neq b_1.$ $C_{b_0b_1}$ gauges the correlation between
photons in modes $b_0$ and $b_1$ \cite{footnote} and depends only
on the statistics of the input light. Note that Eq.
(\ref{two-point-transmission}) is exact and independent of $1/g.$
The correlation function discussed here is fundamentally different
from the one encountered in simpler systems as, e.g., a dualport
beamsplitter, since Eq. (\ref{two-point-transmission}) is averaged
over all ensembles of disorder. To measure the two-point
correlation function generally requires involved coincidence
detection schemes between different output directions from the
medium. However, we find here that the quantum fluctuations of the
total transmission, or equivalently the total reflection, provide
an alternative way of extracting the two-point correlation. The
total transmission and the quantum correlation function are
connected through
\begin{equation} \frac{\overline{\Delta {n}_T^2}}{\left<\hat{n}_a^{in} \right>
} = \frac{\ell}{L} + \frac{\ell^2}{L^2} \left<\hat{n}_a^{in}
\right> \left[ C_{b_0b_1} - 1 \right] , \end{equation}
where again corrections of order $1/g$ have been neglected.

\begin{figure}[t]
\includegraphics[width=\columnwidth]{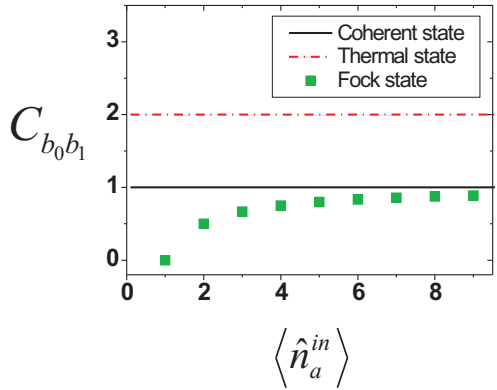}
 \caption{(color online). Two-point correlation function $C_{b_0b_1}$ for three different quantum states of light as a function of the average
 number of photons in the incoming field. The correlation function exhibits either strong correlation (thermal state) or
 anti-correlation (Fock state) relative to the uncorrelated case (coherent state).  }
 \label{Cb0b1}
\end{figure}

Figure \ref{Cb0b1} shows the two-point correlation function for
three different quantum states when varying the average number of
photons in the input state. For a coherent state with an arbitrary
average number of photons the correlation function is unity, i.e.
the spatial correlations vanish. For single mode thermal states,
the spatial correlation function is always equal to $2$, which
means that different parts of the speckle pattern are strongly
correlated. For Fock states, a striking non-classical behavior is
found: the spatial correlation function is reduced below unity,
which means that the spatial directions are anti-correlated. This
corresponds to spatial anti-bunching of light, and the correlation
function $C_{b_0b_1}$ is the spatial counterpart of the
second-order coherence function $g^{(2)}(\tau)$ that is
omnipresent in quantum optics \cite{Loudon}. For a single-photon
Fock state, the correlation function vanishes identically since if
a photon is detected in one channel the probability of detecting
another photon in a different channel is zero. For Fock states
with an increasing number of photons, the correlation function
approaches unity, i.e. the spatial correlations vanish.

\begin{figure}[t]
\includegraphics[width=0.8\columnwidth]{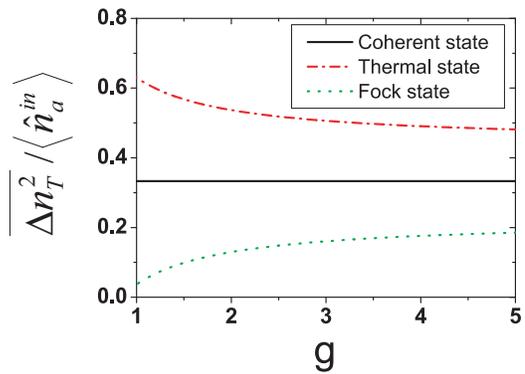}
 \caption{(color online). Fluctuations of the total transmission for a coherent state, a thermal state with $\left<\hat{n}_a^{in}
 \right>=1$, and an arbitrary Fock state
 as a function of the mesoscopic conductance $g$, and for fixed
amount of disorder: $\ell/L=1/3.$ For decreasing
 $g$, the mesoscopic correlations either enhance or suppress
 quantum fluctuations for the thermal state and Fock state, respectively. The fluctuations of the coherent state are
 unaltered for all values of $g$. When $g$ approaches unity, the
 localization regime is entered and the model is invalid.
  }
 \label{g-corr}
\end{figure}

In addition to the spatial correlations also mesoscopic
fluctuations contribute to the quantum fluctuations of the total
transmission, c.f. Eq. (\ref{Total-Fluc}). Figure \ref{g-corr}
displays these fluctuations as a function of the mesoscopic
conductance $g$. For small values of $g$, the mesoscopic
correlations either enhance or reduce the quantum fluctuations for
thermal states and Fock states, respectively. In contrast, the
quantum fluctuations for a coherent state of light are unaltered.
These mesoscopic fluctuations provide another example of
correlations that are present only in a quantum description of
multiple scattering.

It is instructive to compare our findings on photon fluctuations
in diffusive propagation to the transport of electrons through an
elastically scattering disordered conductor. In the diffusive
regime of electron transport, Poissonian electron fluctuations are
reduced, which is a direct consequence of the Fermi-Dirac
statistics of electrons in each individual conduction channel
\cite{Beenakker92,Blanter00}. However, no correlations exist
between different conduction channels and the fluctuations can
simply be added up incoherently. Our work demonstrates that the
dominating correlation in diffusive transport of photons is due to
correlation between different spatial modes that are the optical
analogy to electronic conduction channels, thus pointing out a
fundamental difference between electrons and photons.

We have predicted a spatial correlation present in a quantum
description of multiple scattering of light. The spatial
correlation depends on the quantum state of the light source
illuminating the sample, and can be measured by recording the
quantum fluctuations of the total transmission or reflection from
the sample. Our work shows that multiple scattering of
non-classical light induces stronger correlations than for
classical light. The predictions should be readily observable
experimentally even for moderate amounts of scattering.

This work is part of the research program of the "Stichting voor
Fundamenteel Onderzoek der Materie (FOM)", which is financially
supported by the "Nederlandse Organisatie voor Wetenschappelijk
Onderzoek (NWO)".


\begin{thebibliography}{}


\bibitem{Feng88}
S. Feng, C. Kane, P.A. Lee, and A.D. Stone, Phys. Rev. Lett. {\bf
61}, 834 (1988).

\bibitem{Freund88}
I. Freund, M. Rosenbluh, and S. Feng, Phys. Rev. Lett. {\bf 61},
2328 (1988).

\bibitem{Albada90}
M.P. van Albada, J.F. de Boer, and A. Lagendijk, Phys. Rev. Lett.
{\bf 64}, 2787 (1990).

\bibitem{Scheffold98}
F. Scheffold and G. Maret, Phys. Rev. Lett. {\bf 81}, 5800 (1998).

\bibitem{Shapiro99}
B. Shapiro, Phys. Rev. Lett. {\bf 83}, 4733 (1999).

\bibitem{Martin02}
A. Garcia-Martin, F. Scheffold, M. Nieto-Vesperinas, and J.J.
Saenz, Phys. Rev. Lett. {\bf 88}, 143901 (2002).

\bibitem{Skipetrov04}
S.E. Skipetrov, Phys. Rev. Lett. {\bf 93}, 233901 (2004).

\bibitem{Chabanov00}
A.A. Chabanov, M. Stoytchev, and A.Z. Genack, Nature {\bf 404},
850 (2000).

\bibitem{Patra&Beenakker99}
M. Patra and C.W.J. Beenakker, Phys. Rev. A {\bf 59}, R43 (1999).

\bibitem{Lodahl05}
P. Lodahl and A. Lagendijk, Phys. Rev. Lett. {\bf 94}, 153905
(2005).

\bibitem{Lugiato02}
L.A. Lugiato, A. Gatti, E. Brambilla, J. Opt. B. {\bf 4}, S176
(2002).

\bibitem{Beenakker98}
C.W.J. Beenakker, Phys. Rev. Lett. {\bf 81}, 1829 (1998).

\bibitem{Patra00}
M. Patra and C.W.J. Beenakker, Phys. Rev. A {\bf 61}, 063805
(2000).

\bibitem{commutation relations}
As a direct consequence of the commutation relation for the output
modes, we have the identity
$\left[\hat{a}_{b_0},\hat{a}_{b_1}^{\dagger} \right] = \sum_{a'}
t_{a' b_0} t_{a' b_1}^* + \sum_{b'} r_{b' b_0} r_{b' b_1}^* =
\delta_{b_0,b_1}.$

\bibitem{Berkovits94}
R. Berkovits and S. Feng, Phys. Rep. {\bf 238}, 135 (1994).

\bibitem{Mandel&Wolf}
L. Mandel and E. Wolf, \textit{Optical coherence and quantum
optics} (Cambridge University Press, New York, 1995).

\bibitem{Rossum99}
M.C.W. van Rossum and Th.M. Nieuwenhuizen, Rev. Mod. Phys. {\bf
71}, 313 (1999).

\bibitem{footnote}
We note that the correlation function $C_{b_0b_1}$ defined here is
different from the classical correlation function of relevance for
intensity measurement, as given, e.g., in \cite{Berkovits94}.

\bibitem{Loudon}
R. Loudon, \textit{The quantum theory of light} (Oxford University
Press, New York, 2000).

\bibitem{Beenakker92}
C.W.J. Beenakker and M. B\"{u}ttiker, Phys. Rev. B {\bf 46}, R1889
(1992).

\bibitem{Blanter00}
Ya.M. Blanter and M. B\"{u}ttiker, Phys. Rep. {\bf 336}, 1 (2000).




\newpage

\end{thebibliography}
\end{document}